\documentclass[aps,prapplied,twocolumn,superscriptaddress, noeprint]{revtex4-2}
\usepackage{graphicx}
\usepackage{siunitx}
\usepackage{amsmath}
\usepackage{amssymb}
\usepackage{physics}
\usepackage[T1]{fontenc}

\begin{document}
	
	\title{Intrinsic Noise of the Single-Electron Box}
	
	\author{Laurence Cochrane}
	\email{olc22@cam.ac.uk}
	\affiliation{Nanoscience Centre, Department of Engineering, University of Cambridge, Cambridge CB3 0FF, United Kingdom}
	\affiliation{Quantum Motion, 9 Sterling Way, London N7 9HJ, United Kingdom}
	\author{Ashwin A. Seshia}
	\affiliation{Nanoscience Centre, Department of Engineering, University of Cambridge, Cambridge CB3 0FF, United Kingdom}
	\author{M. Fernando Gonzalez-Zalba}
	\affiliation{Quantum Motion, 9 Sterling Way, London N7 9HJ, United Kingdom}
	
	\date{\today}
	
	\begin{abstract}
		The radio-frequency Single-Electron Box is becoming an attractive charge sensor for semiconductor-based quantum computing devices due to its high sensitivity and small footprint, which facilitates the design of highly connected qubit architectures. However, an understanding of its ultimate sensitivity is missing due to the lack of a noise model. Here, we quantify the intrinsic noise of the Single-Electron Box arising from stochastic cyclic electron tunnelling between a quantum dot and a reservoir driven by a periodic gate voltage. We use both a master equation formalism and Markov Monte Carlo simulations to calculate the gate noise current, and find the noise mechanism can be represented as a cyclostationary process. We consider the implications of this cyclostationary noise on the ultimate sensitivity of Single-Electron Box sensors for fast, high-fidelity readout of spin qubits, in particular evaluating results for radio-frequency reflectometry implementations and the backaction of the sensor on a qubit. Furthermore, we determine the conditions under which the intrinsic noise limit could be measured experimentally and techniques by which the noise can be suppressed to enhance qubit readout fidelity.
	\end{abstract}
	
	\maketitle
	
	\section{Introduction}
	
	Semiconductor-based quantum computing architectures are emerging as promising candidates for large-scale quantum information processing~\cite{chatterjeeSemiconductorQubitsPractice2021}. They offer prospects to leverage industrial fabrication techniques~\cite{maurandCMOSSiliconSpin2016a, zwerverQubitsMadeAdvanced2022} and to be integrated with classical electronics~\cite{ruffinoCryoCMOSChipThat2022a} to scale up current demonstrations of few-qubit processors~\cite{philipsUniversalControlSixqubit2022b, xueQuantumLogicSpin2022}. Fast, high fidelity qubit readout is a critical capability for fault-tolerant operation~\cite{fowlerSurfaceCodesPractical2012}, which in state-of-the-art implementations is provided by charge sensors such as the Single-Electron Transistor (SET)~\cite{keithSingleShotSpinReadout2019, connorsRapidHighFidelitySpinState2020a} probed using radio-frequency (rf) reflectometry techniques~\cite{vigneauProbingQuantumDevices2023}. Recently, the Single-Electron Box (SEB)~\cite{houseHighSensitivityChargeDetection2016, borjansSpinDigitizerHighFidelity2021, niegemannParitySingletTripletHighFidelity2022}, a single-gate quantum dot coupled to a single reservoir and whose energy detuning is periodically varied [Fig. \ref{FIG1}(a)], has attracted increased attention as an alternative sensor to the SET. Demonstrations have achieved comparable readout fidelities to the SET while promising greater scalability and architectural flexibility due to the SEB's reduced footprint and electrode count~\cite{oakesFastHighFidelitySingleShot2023}. 
	
	The intrinsic noise of such single-electron devices, arising from the stochastic nature of electron tunnelling, determines their ultimate sensitivity and as such deserves close attention. In the case of the SET, its shot noise has been well studied using classical `orthodox' theory~\cite{korotkovIntrinsicNoiseSingleelectron1994} and, for low excitation frequencies $\omega_0 \ll 2\pi \left<I\right>/e$, is governed by the well-known Schottky current noise spectral density formula, $S_{I} = 2eF\left<I\right>$, where $e$ is the electronic charge, $\left<I\right>$ is the average current flowing through the device, and the Fano factor $F$ quantifies the correlation between tunnelling events. This expression, derived using a master equation approach~\cite{korotkovIntrinsicNoiseSingleelectron1994} and supported by both Markov Monte Carlo simulations~\cite{ammanChargeEffectTransistor1989} and experimental measurements~\cite{kafanovMeasurementShotNoise2009}, has been applied to analyse the ultimate sensitivity of the rf-SET~\cite{korotkovChargeSensitivityRadio1999, roschierNoisePerformanceRadiofrequency2004, devoretAmplifyingQuantumSignals2000} and its measurement backaction~\cite{aassimeRadioFrequencySingleElectronTransistor2001}. Recently, rf-SET based single shot spin readout has approached the theoretical shot noise limit to within an order of magnitude~\cite{keithSingleShotSpinReadout2019}, with further improvements expected by using quantum-limited parametric amplification.  In the case of the SEB, however, a description of its intrinsic noise limit, analogous to that known for the SET, is missing. 
	
	Previous theoretical and experimental studies of the noise associated with cyclic electron tunnelling in the SEB, including investigations of mesoscopic statistical mechanics~\cite{averinStatisticsDissipatedEnergy2011} and quantum thermodynamics~\cite{pekolaQuantumThermodynamicsElectronic2015, koskiExperimentalRealizationSzilard2014, koskiExperimentalRealizationSzilard2014}, have focused on the limit of quasi-adiabatic driving conditions.
    However, as we shall demonstrate, the high-frequency excitation used in rf-reflectometry results in a current noise spectral density that is itself a periodic function of time $S_I(t, \omega)$, a property that must be taken into account for an accurate evaluation of the readout sensitivity. 
    
    Inspiration to quantify $S_I(t, \omega)$ can be drawn from studies of the accuracy of single electron emitters used as ultra-precise current sources \cite{albertAccuracyQuantumCapacitor2010, maheCurrentCorrelationsOndemand2010, parmentierCurrentNoiseSpectrum2012}, for instance a mesoscopic capacitor excited by a high-frequency square wave to produce a quantized current. Experimental measurements of the time-averaged noise spectrum $S_I^0(\omega) = \overline{\langle S_I(t, \omega) \rangle}^t$ can demonstrate the accuracy of the source in emitting exactly one electron per cycle, analogous to interferometric methods used in the characterisation of single photon emitters \cite{maheCurrentCorrelationsOndemand2010}. The experimental dependence of the noise power on electron tunnelling rate $\Gamma_0$ and excitation frequency $\omega_0$ can be well modelled by both a conceptually simple semiclassical model \cite{albertAccuracyQuantumCapacitor2010} and a more complete account using Floquet scattering theory \cite{parmentierCurrentNoiseSpectrum2012}, with simulations based on a tight binding chain model \cite{jonckheereRealtimeSimulationFinitefrequency2012} supporting this understanding. 
	
    However, as we shall show, the time-averaged noise alone is insufficient to characterize the SEB, since correlations in the time-varying noise spectrum can have a significant effect on the signal-to-noise ratio (SNR) after synchronous demodulation. To quantify these correlations, we use the Fourier components $S_I^{n\omega_0}(\omega) = \int S_I(t, \omega) e^{-jn\omega_0t}\,dt$ of the periodically varying $S_I(t, \omega)$, which we call ``spectral correlations functions" following the nomenclature associated to cyclostationary processes~\cite{gardnerCyclostationarityHalfCentury2006}. We note that, in the context of single-electron emitters, these spectral correlations have been termed ``noise harmonics"~\cite{dittmannFinitefrequencyNoiseInteracting2018} or ``photon-induced noise"~\cite{moskaletsNoiseSingleelectronEmitter2013} and have been used to investigate signatures of the underlying dynamics of single-electron emitters.
    
	More particularly, in this Article, we present an analysis of stochastic tunnelling in the SEB under periodic excitation and provide  results relevant to the practical implementation of high fidelity spin qubit readout using rf-reflectometry. We use a semiclassical master equation approach to represent periodically driven tunnelling as a cyclostationary process~\cite{gardnerCyclostationarityHalfCentury2006} and complete the SEB's small-signal equivalent circuit representation (consisting of quantum capacitance and Sisyphus resistance~\cite{gonzalez-zalbaProbingLimitsGatebased2015}) with a noise current generator $I_N$ [Fig. \ref{FIG1}(b)]. We support our analytical expressions for the noise spectrum with Monte Carlo simulations of the underlying time-inhomogeneous Markov process. We study the intrinsic noise spectral density as a function of the SEB parameters and driving conditions, and quantify the baseband noise in rf-reflectometry readout and the backaction induced by the SEB on a nearby qubit. We demonstrate correlations in the noise spectrum lead to noise enhancement or suppression that depends on the demodulation quadrature, a signature of the SEB's intrinsic noise that can be experimentally verified and used to optimise readout fidelities.
	
	\section{Gate Noise Current}
	
	\begin{figure}
		\includegraphics{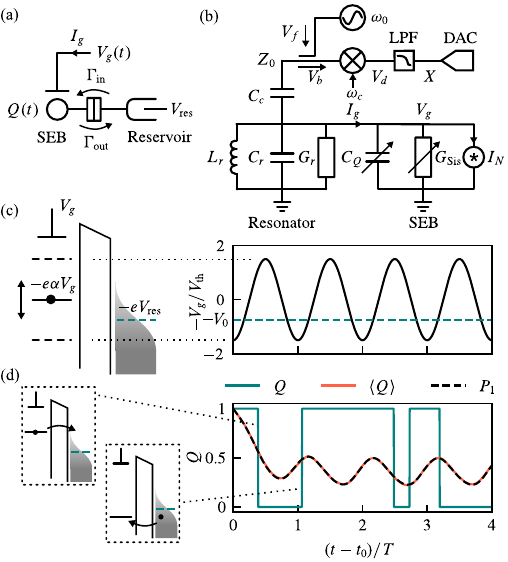}
		\caption{(a)~Schematic of the Single-Electron Box (SEB). (b)~Small-signal equivalent circuit representation of the SEB embedded in a generalised rf-reflectometry setup with a capacitively coupled parallel LC resonator and synchronous demodulation. (c)~Periodic gate excitation signal and (d)~corresponding SEB charge response for an initial dot occupation condition $Q(t_0) = 1$: instance of the stochastic process $Q(t)$ generated by Monte Carlo simulation (solid green); the ensemble expectation $\left<Q(t)\right>$ calculated from the mean of $10^4$ simulated traces (solid orange); and the solution $P_1(t,t_0)$ to the master equation (dashed black). The energy diagrams illustrate in panel (c) the periodic driving of the dot level with respect to the reservoir Fermi energy and in panel (d) the discrete tunnelling in and out events that constitute the stochastic $Q(t)$ time-trace.}
		\label{FIG1}
	\end{figure}

	The SEB consists of a driven, gate-controlled semiconductor quantum dot coupled via a tunnel barrier to a reservoir [Fig.~\ref{FIG1}(a)]. When embedded in a resonant circuit [Fig.~\ref{FIG1}(b)], it modulates the resonant frequency and quality factor through its small-signal gate admittance; this can be inferred from changes in the complex rf reflection coefficient, a scheme termed rf-reflectometry~\cite{vigneauProbingQuantumDevices2023}. Considering in our analysis a single  quantum dot charging level, we model dot-to-reservoir tunnelling in the thermally-broadened regime ($\hbar\Gamma_0 < k_\text{B}T_e$ for a reservoir electron temperature $T_e$ and a maximum tunnel rate $\Gamma_0$) as a Poissonian process with rate parameter~\cite{vigneauProbingQuantumDevices2023}
	\begin{equation}
		\Gamma_\text{in(out)} = \frac{\Gamma_0}{1+\exp{\left(\mp\frac{V_g-V_0}{V_\text{th}}\right)}},
		\label{rate}
	\end{equation}
	where $V_0$ is the voltage that aligns the dot level with the reservoir's Fermi energy level, and $V_\text{th} = k_BT_e/e\alpha$. Here, $\alpha = C_G/C_\Sigma$ is the gate lever arm with $C_\Sigma = (C_G+C_S)$, the sum of dot-gate capacitance $C_G$ and tunnel barrier capacitance $C_S$. The stochastic evolution of the dot electron occupation number $Q(t)$ is described by the master equation~\cite{vigneauProbingQuantumDevices2023}
	\begin{equation}
		\dot{P_1}(t,t_0) + \Gamma_0 P_1(t,t_0) = \Gamma_\text{in}(t),
	\end{equation}
	where $P_1(t,t_0) \equiv \Pr{\left(Q(t) = 1 | Q(t_0) = 1\right)}$ for $ t\geq t_0$, i.e. the probability of finding the quantum dot occupied at time $t$ given it was occupied at time $t_0$. The solution for the initial condition $P_1(t_0,t_0) = 1$ can be written as
	\begin{equation}
		P_1(t,t_0) = P_\text{SS}(t) + e^{-\Gamma_0(t-t_0)}\left[1-P_\text{SS}(t_0)\right].
	\end{equation}
	where $P_\text{SS}(t)$ is the steady state solution reached at times $t-t_0 \gg 1/\Gamma_0$:
	\begin{equation}
		P_\text{SS}(t) = e^{-\Gamma_0 t} \int e^{\Gamma_0 t} \Gamma_\text{in}(t)\, dt.
		\label{pss}
	\end{equation}
	Under a small-signal gate excitation $V_g = V_\text{dev}\cos \omega_0t$ with $ V_\text{dev}/V_\text{th} \lesssim 1$ [Fig.~\ref{FIG1}(c)], Eq.~\eqref{pss} may be analytically evaluated by linearising Eq.~\eqref{rate} as 
	\begin{equation}
		\Gamma_\text{in} (V_g) \approx \frac{\Gamma_0}{1+ \exp(V_0/V_\text{th})} + \frac{\Gamma_0}{4\cosh^2(V_0/2V_\text{th})}\frac{V_g}{V_\text{th}}.
	\end{equation} 
    We arrive at an expression of the steady state probability:
	\begin{equation}
		P_\text{SS}(t) = P_\text{th} + \frac{V_\text{dev}}{e\alpha\omega_0}\big(G_\text{Sis} \sin{\omega_0 t} + \omega_0 C_Q\cos{\omega_0 t}\big)
		\label{Eq:master_sol}
	\end{equation}
    allowing the SEB to be represented by a small-signal admittance formed of a quantum capacitance and a Sisyphus conductance, $Y_Q = I_g(\omega_0)/V_g(\omega_0) = j\omega_0C_Q + G_\text{Sis}$, through which an oscillatory gate current $I_g = e\alpha \dot{P}_\text{SS}$ flows, as shown in the equivalent circuit in Fig.~\ref{FIG1}(b). Here,
	\begin{equation}
		\begin{split}
			P_\text{th} &= \frac{1}{1 + \exp(V_0/V_\text{th})}\\
			C_Q & = \frac{e\alpha}{4V_\text{th}\cosh^2(V_0/2V_\text{th})}\frac{\Gamma_0^2}{\Gamma_0^2 + \omega_0^2}\\
			G_\text{Sis} &= \frac{e\alpha}{4V_\text{th}\cosh^2(V_0/2V_\text{th})}\frac{\omega_0^2\Gamma_0}{\Gamma_0^2 + \omega_0^2}.
		\end{split}
        \label{quant_admit}
	\end{equation}
 
	For a large gate excitation $V_g = V_\text{dev}\cos \omega_0t$ with $V_\text{dev}/V_\text{th} \gg 1$, the SEB response becomes nonlinear as $\Gamma_\text{in}(t)$ in Eq.~\eqref{rate} approaches a square wave of amplitude $\Gamma_0$ and period $T = 2\pi/\omega_0$. A piecewise evaluation of Eq.~\eqref{pss} gives
	\begin{equation}
		P_\text{SS}(t) = \begin{cases}
			1 - \beta e^{-\Gamma_0 \left(t - \left(k - \frac{1}{4}\right)T\right)}, & k-\frac{1}{4} \leq \frac{t}{T} < k+\frac{1}{4} \\
			\beta e^{-\Gamma_0 \left(t - \left(k + \frac{1}{4}\right)T\right)}, &  k+\frac{1}{4} \leq \frac{t}{T} < k+\frac{3}{4} 
		\end{cases}
		\label{Pss_Square}
	\end{equation}
	where $ \beta = \left(1+\exp\left(-\Gamma_0T/2\right)\right)^{-1}$ and $k$ is an integer.
	
	In order to calculate the noise current, we must consider the underlying random process, an ensemble of time-traces $Q(t)$ of which $P_1(t, t_0)$ is the expectation value [Fig.~\ref{FIG1}(d)]. The master equation can be interpreted as the Kolmogorov Forward Equation for the time-inhomogeneous Markov process
	\begin{equation}
		\Pr{\left(Q(t+\delta t) = 1\right)} = 
		\begin{cases}
			\Gamma_\text{in}(t)\delta t,\ &Q(t) = 0\\
			1 - \Gamma_\text{out}(t)\delta t,\ &Q(t) = 1.
		\end{cases}
		\label{MMC}
	\end{equation}
	Implementing this as a Markov Monte Carlo (MMC) simulation allows us to generate instances of the random process $Q(t)$ and verify the ensemble expectation $\left<Q(t)\right>$ tends to $P_1(t)$ [Fig.~\ref{FIG1}(d)].
	
\begin{figure}[t]
		\includegraphics{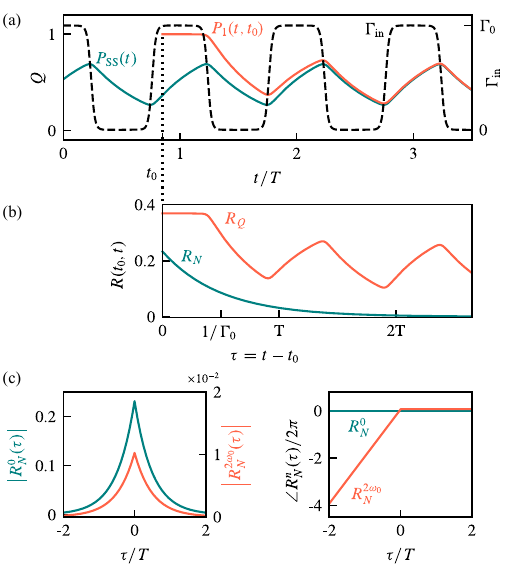}
		\caption{Derivation of cyclic autocorrelation functions. (a)~Steady-state $P_\text{SS}(t)$ (green) and conditional $P_1(t, t_0)$ (orange) SEB occupation probabilities in response to a driving term $\Gamma_\text{in}(t)$ (dashed black) in large excitation regime $V_\text{dev}/V_\text{th} = 10$. The charge autocorrelation $R_Q(t, t_0)$ is given by their product $P_\text{SS}(t)P_1(t, t_0)$ [panel (b) - orange]; the noise autocorrelation $R_N$ (green) is isolated by subtraction of the mean. (c) Magnitude and phase of the fundamental and second harmonic cyclic autocorrelation functions (Fourier components of $R_N$), the orders that contribute to the measured noise in rf-reflectometry.}
		\label{FIG2}
	\end{figure}

	The noise spectrum is given by the Fourier transform of the random process's autocorrelation function, $R_Q(t_0, t) = \mathbb{E} \left[Q(t_0) Q(t)\right]$, calculated from the joint probability that the dot is occupied at both times $t_0$ and $t$.  For $t \geq t_0$, this can be understood in terms of conditional probabilities and expressed in terms of the solution to the master equation, as illustrated in Fig. \ref{FIG2}(a) and (b):
	\begin{equation}
		\begin{split}
			R_Q(t_0, t) &= \Pr \left(Q(t_0) = 1 \right) \Pr \left( Q(t) = 1 | Q(t_0) = 1 \right) \\
			&= P_\text{SS}(t_0)P_1(t, t_0)
		\end{split}
		\label{ACF}
	\end{equation}
	As both the ensemble mean, $P_\text{SS}(t)$, and autocorrelation, $R_Q(t_0, t)$, are periodic ($R_Q(t_0, t) = R_Q(t_0 + kT, t +kT)$ for all integers $k$), $Q(t)$ conforms to the definition of a cyclostationary process~\cite{gardnerCyclostationarityHalfCentury2006}.	We can isolate the noise fluctuations by defining the random process $N(t) = Q(t) - \left<Q(t)\right>$ with autocorrelation $R_N(t_0, t) =  P_\text{SS}(t_0)\left(P_1(t_0, t)-P_\text{SS}(t)\right)$ [Fig. \ref{FIG2}(b)]. Under the transformation $\tau = t- t_0$, this can be expressed as a Fourier series of cyclic autocorrelation functions $R^{n\omega_0}_{N}(\tau)$,
	\begin{equation}
		R_{N}(t,\tau) = \sum_{n = -\infty}^{+\infty} R^{n\omega_0}_{N}(\tau) e^{jn\omega_0 t},
		\label{Fourier}
	\end{equation}
	plotted in Fig.~\ref{FIG2}(c) for $n = 0$ and 2 which are the orders that, as we derive below, contribute to the noise in rf-reflectometry. The timescale over which the noise autocorrelation decays is set by the inverse of the tunnel rate $\Gamma_0$.
	
	The Fourier transform of Eq.~\eqref{Fourier} gives the charge noise spectrum $S_N(t,\omega)$ in terms of a series of spectral correlation functions, $S_N^{n\omega_0}(\omega)$. These are the equivalent of the noise harmonics in \cite{dittmannFinitefrequencyNoiseInteracting2018} and can be interpreted as the time-averaged correlation between frequency-shifted copies of the spectrum, $S_N^\alpha(\omega) = \overline{\left<S_N(t,\omega)S_N(t,\omega+\alpha) \right>}^t$~\cite{gardnerCyclostationarityHalfCentury2006}; the zeroth order is the time-averaged noise used to characterise the accuracy of single electron emitters \cite{parmentierCurrentNoiseSpectrum2012}. The transformation from dot charge fluctuations $eN(t)$ to gate current noise is given by the Norton equivalent current $I_N = (e\alpha) dN/dt$ in parallel with $C_S$, which we treat as negligible in comparison to external parasitic capacitances. In terms of the spectral correlation functions, the transformation from charge to current noise is implemented as \cite{gardnerCyclostationarityHalfCentury2006}
	\begin{equation}
		S_I^{n\omega_0}[\omega] = (e\alpha)^2\omega(\omega-n\omega_0)S_N^{n\omega_0}[\omega].
		\label{NtoI}
	\end{equation}
	In the small-excitation limit, the $n = 0$ and $2$ orders evaluate to (see appendix A)
	\begin{equation}
		S_I^0(\omega) = \left[\frac{(e\alpha)^2}{2\cosh^2(V_0/2V_\text{th})}- \left(\frac{V_\text{dev}|Y_Q|}{\omega_0}\right)^2  \right] \frac{\Gamma_0\omega^2}{\Gamma_0^2 + \omega^2}
		\label{Eq:SI0_SS}
	\end{equation}
	\begin{equation}
		\begin{split}
			S_I^{2\omega_0} = &-\left(\frac{V_\text{dev}Y_Q}{2\omega_0}\right)^2 \omega(\omega-2\omega_0) \\
							& \times\left[\frac{1}{\Gamma_0 + j\omega} + 	\frac{1}{\Gamma_0 - j(\omega-2\omega_0)}\right],
		\end{split}
	\end{equation}
	whereas for large excitation, the corresponding expressions are:
	\begin{equation}
		S^0_I(\omega) = (e\alpha)^2\frac{\omega_0}{\pi}\tanh\left(\frac{\pi\Gamma_0}{2\omega_0}\right)\frac{\omega^2}{\Gamma_0^2 + \omega^2}
	\end{equation}
	\begin{equation}
		\begin{split}
			S_I^{2\omega_0}(\omega) = &(e\alpha)^2 \frac{\omega_0}{\pi}\tanh\left(\frac{\pi\Gamma_0}{2\omega_0}\right) \omega(\omega-2\omega_0) \\
			&\times\left[\frac{1}{\Gamma_0+2j\omega_0} - \frac{1}{2(\Gamma_0+j\omega_0)}\right]\\
			&\times\left[\frac{1}{\Gamma_0 + j\omega} + \frac{1}{\Gamma_0 - j(\omega-2\omega_0)}\right].
		\end{split}
	\end{equation}
	It is worth noting that in the zero-excitation limit of Eq.~\eqref{Eq:SI0_SS}, the noise spectral density at $\omega_0$ tends to the Johnson noise value predicted by the fluctuation-dissipation theorem using the SEB's Sisyphus conductance: $S_I^0(\omega_0) = 2k_\text{B}T_e G_\text{Sis}$. Conversely, this does not hold in the strongly-driven case, where the out-of-equilibrium system dynamics do not obey detailed balance. 
    
	\begin{figure*}
		\includegraphics{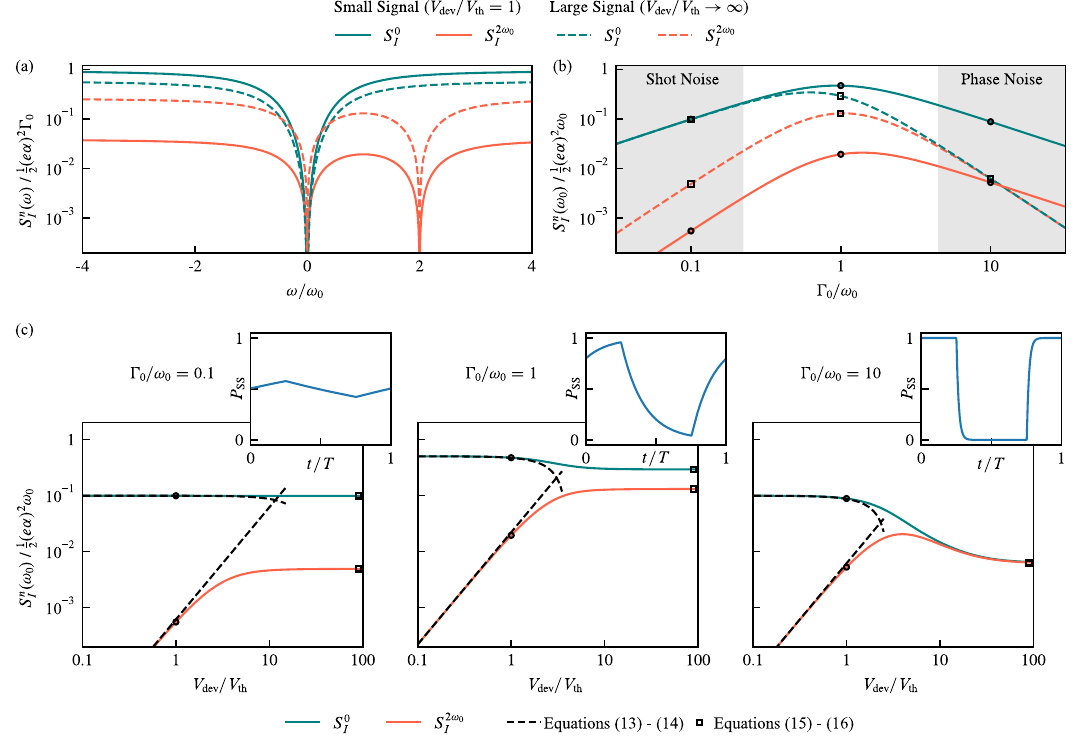}
		\caption{Gate current noise of the SEB quantified by the magnitude of the spectral correlation functions $S_I^n$, calculated from the numerical solution to the master equation $P_\text{SS}(t)$.
			(a)~Fundamental (green) and second order ($2\omega_0$ - orange) spectra for  $\Gamma_0 = \omega_0$ in the small excitation ($V_\text{dev}/V_\text{th} = 1$ - solid lines) and large excitation (dashed lines) regimes. 
			(b)~Dependence of the spectral density at $\omega_0$ on tunnel rate $\Gamma_0$ across the shot and phase noise (``quantum jitter") regimes (small to large $\Gamma_0/\omega_0$) for small (solid) and large (dashed) excitations. The system response in the shot, intermediate and phase noise regimes is illustrated in panels (c)-(e), including the dependence of the spectral density at $\omega_0$ on excitation amplitude $V_\text{dev}$ (numerical calculations - green/orange; analytical expressions -dashed lines, square markers) and the large-excitation steady-state dot charge response $P_\text{SS}$ (inset). The circular (small excitation) and square (large excitation) markers indicated corresponding points in panels (b) and (c).}
		\label{FIG3}
	\end{figure*}
	
	We now evaluate the spectral correlation functions and explain their physical meaning in different limits. Figure~\ref{FIG3} illustrates the dependence of the spectral correlation densities on frequency, tunnel rate and excitation amplitude, and compares the above analytical expressions to the numerical evaluation from the exact solution to the master equation $P_\text{SS}(t)$ propagated through equations \eqref{ACF} to \eqref{NtoI}. In panel~\ref{FIG3}(a), we show the fundamental ($S_I^0$) and second order ($S_I^{2\omega_0}$) spectral correlation functions in the small-excitation (solid lines) and large-excitation  regimes (dashed lines). The spectral densities tend to 0 at zero frequency, consistent with the zero DC current imposed by the cyclic tunnelling, while tending to a flat (white noise) spectrum in the high-frequency limit. Around the driving frequency $\omega_0$, the portion of the spectrum which we shall show contributes to the noise in rf-reflectometry applications, there is a local maximum in the degree of correlation, as quantified by $S_I^{2\omega_0}$. 
	
	We can identify two contrasting regimes of operation for the intrinsic noise mechanism, the ``shot" and ``phase" noise regimes occurring in the limit of slow or fast tunnel rates compared to the excitation frequency respectively [Fig. \ref{FIG3}(b)]. In the limit of small $\Gamma_0/\omega_0$ - i.e. slow tunnelling or fast excitation -  tunnelling occurs more rarely than each excitation period. The independent events result in shot-like noise with time-averaged spectral density $S_I^0(\omega)$ proportional to $\Gamma_0$, tending to the Schottky limit of $2e\langle I \rangle$ for a mean current $\langle I \rangle = e\Gamma_0/2$. Conversely, in the limit of large $\Gamma_0/\omega_0$ (slow excitation), tunnelling occurs each half-period of the drive, with noise arising from the exponentially distributed random times at which the tunnelling event occurs within the period. This can be understood as phase noise or ``quantum jitter" \cite{maheCurrentCorrelationsOndemand2010} and results in a time-averaged spectral density proportional to $\omega_0^2/\Gamma_0$ (in the small-signal limit) or $\omega_0^3/\Gamma_0^2$ (for large excitations), as can be seen from the slopes of the $n = 0$ orders at large $\Gamma_0/\omega_0$ in Fig. \ref{FIG3}(b). The noise is maximised in the intermediate regime (at $\Gamma_0/\omega_0 = 1$ for small excitations), which also corresponds to the conditions for maximum Sisyphus conductance (see Eq.~\eqref{quant_admit}).
	
	In Fig.~\ref{FIG3}(c), we study the dependence of the spectral correlation densities at $\omega_0$ on the excitation amplitude $V_\text{dev}/V_\text{th}$ at three tunnel rates, $\Gamma_0/\omega_0 = 0.1$, 1 and 10, corresponding to the shot, intermediate and phase noise regimes respectively. We find good agreement between the numerical and small-signal analytical expressions for $V_\text{dev}/V_\text{th} \lesssim 1$, and between the numerical and large-excitation expressions for $V_\text{dev}/V_\text{th} \gtrsim 50$. In each case, we observe two trends: the average noise power $S_I^0(\omega_0)$ decreases at large excitation amplitudes, while the ``spectral coherence"~\cite{gardnerUnifyingViewCoherence1992}, i.e. the ratio spectral correlation densities $\rho = \left|S_I^{2\omega_0}(\omega_0)/S_I^0(\omega_0)\right|$, increases with excitation amplitude. The decrease in $S_I^0(\omega_0)$ can be intuitively explained by considering $P_\text{SS}(t)$, which deviates further from 0.5 at large excitation amplitudes, corresponding to a reduced variance in $Q(t)$ and hence reduced noise current. As we shall discuss in the next section, the coherence plays an important role in noise enhancement or suppression during demodulation. We find this is maximised in the phase noise regime; again, the master equation solutions $P_\text{SS}(t)$ in the high $V_\text{dev}/V_\text{th}$ limit (see Fig.~\ref{FIG3}(c) insets) can help give an intuitive understanding. At large $\Gamma_0/\omega_0$, $P_\text{SS}(t)$ approaches a square wave, indicating two near-deterministic tunnelling events per cycle; these temporal correlations between cycles correspond to strong cyclostationary spectral correlations. By contrast, at small $\Gamma_0/\omega_0$, the resultant $P_\text{SS}(t) \approx 0.5$ means $Q(t)$ approaches a ``random telegraph" signal which does not exhibit any such correlations. 
 
	\section{Intrinsic Signal-to-Noise Ratio}
	
	\begin{figure*}
		\includegraphics{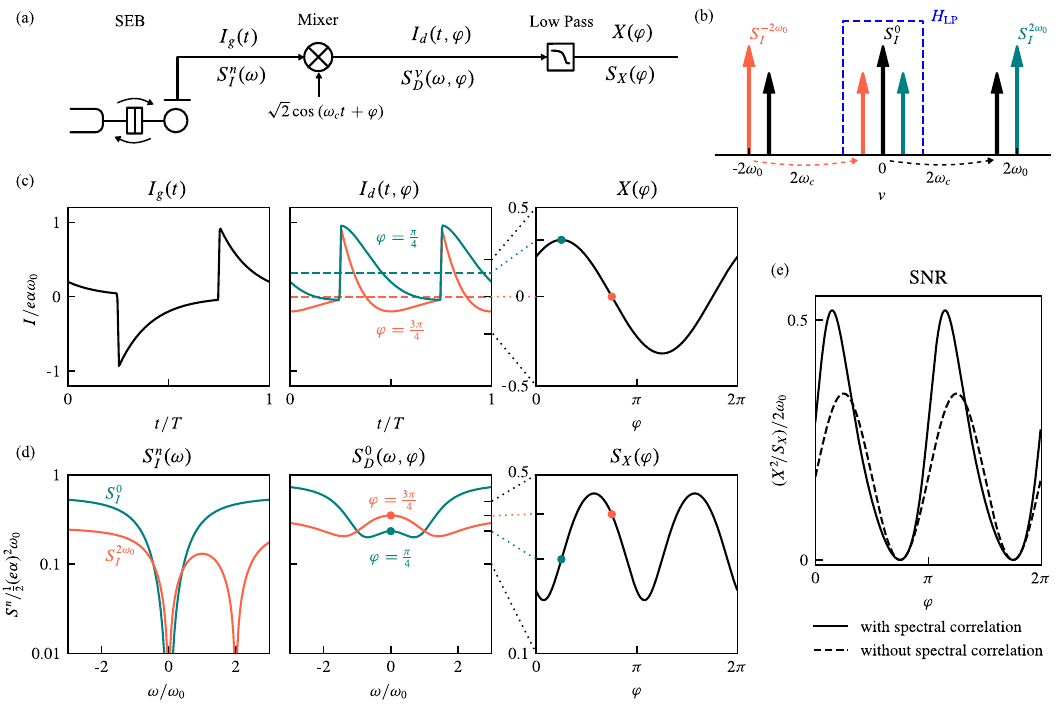}
		\caption{Intrinsic Signal-to-Noise Ratio of the SEB's gate current. (a)~The synchronous demodulation scheme applied to the gate current $I_g$, consisting of downmixing with a local oscillator of frequency $\omega_c$ and quadrature phase $\varphi$ followed by low pass filtering with transfer function $H_\text{LP}$. The signals $I_g$, $I_d$ and $X$ at each stage of demodulation and the corresponding noise spectra $S_I$, $S_D$ and $S_X$ are illustrated in panels (c) and (d) respectively. (b)~Generation of the cyclostationary orders $\nu$ of the mixer output ($I_d$) from the input spectral correlation functions $S_I^{n\omega_0}$ and local oscillator frequency $\omega_c$ ($S_I^{\pm\omega_0}$ orders omitted for clarity); the pass-band of the image rejection filter with transfer function $H_\text{LP}$ is shown in blue. (c)~Example signal waveforms and (d)~noise spectra in the large-excitation limit with $\omega_0 = \Gamma_0$ and $\omega_c = \omega_0$. In the middle panel two demodulation quadratures $\varphi = \pi/4$ and $3\pi/4$ corresponding to the minimum and maximum demodulated signal are shown. The output of the low pass filter $X$ is the DC component of the downmixed waveform $I_d$ (dashed lines) (e)~Resultant SNR as a function of demodulation quadrature $\varphi$, both with and without the effect of the cyclostationary spectral correlation $S_I^{2\omega_0}$.}
		\label{FIG4}
	\end{figure*}
	
	Given the periodicity of the gate current of the SEB and the cyclostationary properties of the SEB's noise, the measurement signal-to-noise ratio (SNR) can only be defined in the context of synchronous demodulation (lock-in detection), as is used in homodyne rf-reflectometry. In this section, we consider the idealised case of direct demodulation of the gate current $I_g$ emitted by the SEB to determine the intrinsic SNR of the sensor without external factors such as the electrical resonator or amplifier noise; the effect of the resonator will be explored in section V. 
	
	The signal processing chain applied to the gate current $I_g(t)$, depicted in Fig.~\ref{FIG4}(a), consists of downmixing by a local oscillator $\sqrt{2}\cos(\omega_ct+\varphi)$ with output $I_d(t)$ and image rejection through a low pass filter $H_\text{LP}(\omega)$ to give a baseband signal $X(t, \varphi)$ which depends on the demodulation quadrature $\varphi$. Downmixing produces a new cyclostationary signal $I_d(t)$ with cyclostationary orders $\nu \in \left\{n\omega_0 \cup n\omega_0\pm 2\omega_c\right\}$, illustrated schematically in Fig.~\ref{FIG4}(b); we can write down the corresponding spectral density $S_{D}^\nu(\omega, \varphi)$ as~\cite{gardnerCyclostationarityHalfCentury2006}:
	\begin{equation}
		\begin{split}
			S_{D}^\nu(\omega, \varphi) = &\frac{1}{2}\Big[S_I^\nu(\omega-\omega_c) + S_I^\nu(\omega+\omega_c) + \\ 
			&S_I^{\nu+2\omega_c}(\omega+\omega_c)e^{-j2\varphi} +     S_I^{\nu-2\omega_c}(\omega-\omega_c)e^{j2\varphi}\Big].
		\end{split}
		\label{mix}
	\end{equation}

	We are interested in the experimentally relevant case of homodyne detection with $\omega_c = \omega_0$, in which $I_d(t, \varphi)$ becomes a DC signal $X(\varphi)$ after the image rejection filter $H_\text{LP}$. In this case, the cyclostationary orders of $S_D^\nu$ become $\nu \in {n\omega_0}$; given a filter cut-off frequency $\omega_\text{LP} < \omega_0/2$, the filter output $X$ is then a wide-sense stationary process with spectral density $S_X(\omega, \varphi) = S_D^0(\omega, \varphi)$ for $|\omega|<\omega_{LP}$, since cyclostationary spectral correlations at frequencies separated by $\nu = n\omega_0$ cannot be supported by a signal band-limited to $|\omega|<\omega_0/2$~\cite{gardnerCyclostationarityHalfCentury2006}. Evaluating  Eq.~\eqref{mix} for $\omega_c = \omega_0$, we find the low-frequency spectral density $S_X(\varphi) \equiv S_X(0, \varphi)$ after demodulation is given by
    \begin{equation}
        S_X(\varphi) = S_D^0(0, \varphi) = S_I^0(\omega_0) + \left|S_I^{2\omega_0}(\omega_0)\right| \cos(2\varphi + \theta),
        \label{demodnoise}
    \end{equation}
    where $\theta = \arg{\left(S_I^{2\omega_0}(\omega_0)\right)}$. Equation~\eqref{demodnoise} demonstrates a key signature of the SEB's intrinsic noise, one of the main results of our Article: cyclostationary spectral correlations, in particular the orders $n = 0$ and $\pm2$ of $S_I^{n\omega_0}$, result in enhanced or suppressed noise depending on the local oscillator phase $\varphi$. For convenience, we notate this result as $S_X(\varphi) = S_X^0 + \Delta S_X \cos(2\varphi + \theta)$ where $S_X^0 = S_I^0(\omega_0)$ and $\Delta S_X = \left|S_I^{2\omega_0}(\omega_0)\right|$. Under conditions (as explored in section II) where the coherence $\rho = \left|S_I^{2\omega_0}(\omega_0)/S_I^0(\omega_0)\right| = \Delta S_X/S_X^0$ tends to 1, we can achieve near-complete suppression of the intrinsic noise in the demodulation quadrature $\varphi = (\pi - \theta)/2$.
    
	Figures \ref{FIG4}(c) and (d) illustrate step-by-step the transformations described above, depicting both the signals and noise spectral densities at each stage from the device gate current and mixer output to the final low pass filtered baseband signal. The values are calculated for $\Gamma_0 = \omega_0$ in the large-excitation limit, conditions chosen to maximise both the signal $I_g$ and the noise $S_I^0$ while also demonstrating the effect of a high degree of spectral coherence ($\rho = 0.45$). In the middle panel, two orthogonal local oscillator phases ($\varphi = \pi/4$ and $3\pi/4$) are shown, corresponding to standard IQ demodulation with the quadrature axes chosen to align with maximum and zero demodulator output signal respectively. The phase-dependence of the SEB's intrinsic noise, a purely cyclostationary effect that contrasts with the phase-independence of uncorrelated noise processes such as amplifier thermal noise, can clearly be seen in the differences between the spectra $S_D^0$ for the two quadratures [middle panel of Fig.~\ref{FIG4}(d)] and in the sinusoidal variation of $S_X(\varphi)$ [right panel of Fig.~\ref{FIG4}(d)]. Note that the noise peaks and troughs in the right panel do not necessarily correspond to the quadratures of maximum and zero signal; it follows that the optimum demodulator phase that maximises the signal-to-noise ratio $\text{SNR} = X^2/(2S_X \Delta f)$ (where $\Delta f = 1/2\tau_\text{int}$ is the noise bandwidth corresponding to an integration time $\tau_\text{int}$~\cite{vigneauProbingQuantumDevices2023}) is not the same as the phase that maximises the signal, as can be seen in Fig.~\ref{FIG4}(e) (solid line). By contrast, omitting the effect of the cyclostationary spectral correlation $S_I^{2\omega_0}$ and only accounting for the time-averaged noise $S_I^0$ would predict an SNR that follows the phase dependence of the signal magnitude $X^2$, as shown by the dashed line in Fig.~\ref{FIG4}(e). From the maximum of the SNR plot we can estimate the ideal sensitivity of the SEB, as characterised by the minimum integration time required to achieve a power SNR = 1~\cite{oakesFastHighFidelitySingleShot2023}. For the SEB and excitation parameters used in this example (giving a maximum $X^2/S_X \approx \omega_0$), we find an integration time shorter than a single period of the drive, $\tau_\text{int} = 1/\omega_0 = T/2\pi$, is sufficient to distinguish between SEB states on and off Coulomb blockade. However, a more extensive study over the parameter space, including the SEB response to small changes in detuning $V_0$, will be required to determine the ultimate performance limits of the SEB and draw a comparison to those of the SET. 
	
	\section{SEB-Induced Backaction}
	
	Coupling the SEB to a qubit can lead to dephasing and relaxation, with this backaction mediated by fluctuations of the voltage  $V_\text{QD}$ on the SEB quantum dot both during and in between measurement operations. The stochastic dot charge $Q(t)$ and the periodic gate drive $V_g(t)$ both contribute to to the fluctuations $V_\text{QD}(t) = -eQ(t)/C_\Sigma + \alpha V_g(t)$, which result in a noise term $S_V(\omega)$ arising from the charge noise autocorrelation $R_N$, as well as a harmonic driving term due to the periodic $V_g(t)$ and $P_\text{SS}(t)$ (see appendix B):
	\begin{equation}
		S_V(\omega) = \left(\frac{e}{C_\Sigma}\right)^2S_N^0(\omega) = \left(\frac{1}{\omega C_G}\right)^2S_I^0(\omega)
	\end{equation}

	The backaction-induced relaxation rate $\Gamma_1$ of a qubit $\ket{0}$, $\ket{1}$ with Hamiltonian $H_Q$ is proportional to the noise spectral density at the qubit frequency $\omega_q$ and are given by the well-known formula \cite{vigneauProbingQuantumDevices2023, aassimeRadiofrequencySingleelectronTransistor2001b}:
	\begin{equation}
		\Gamma_1 = \frac{1}{T_1} = \frac{1}{2\hbar^2}\left| \bra{0}\frac{\partial H_Q}{\partial V_\text{QD}} \ket{1}\right|^2 S_V(\omega_q).
	\end{equation}
	The harmonic component of $V_\text{QD}$ can be separately considered as driving Rabi oscillations (negligible if $\omega_q \neq n\omega_0$). On the other hand, the dephasing rate $\Gamma_{\phi}$ can be expressed in terms of the low-frequency noise \cite{vigneauProbingQuantumDevices2023}
	\begin{equation}
		\Gamma_{\phi} = \frac{1}{T_2^*} = \frac{1}{4} \left(\frac{\partial \omega_q}{\partial V_\text{QD}}\right)^2 S_V(0),
	\end{equation}
	while a constant shift in the qubit frequency due to the DC term of $V_\text{QD}$, $eP_\text{th}/C_\Sigma$, may be accounted for separately. 
	
	We consider the example of a double quantum dot charge qubit with interdot tunnel coupling $\Delta_c$, operated at interdot energy detuning $\varepsilon_0$ and capacitively coupled with lever arm $\kappa$ to the SEB. Given the Hamiltonian of the system $H_\text{DQD} = \frac{1}{2}(\varepsilon\sigma_z + \Delta_c\sigma_x)$ and the total detuning $\varepsilon = \varepsilon_0 + \kappa V_\text{QD}$, we find the SEB-induced realxation and dephasing rates:
	\begin{equation}
		\Gamma_1  = \frac{1}{2\hbar^2}\left(\frac{e\kappa \Delta_c}{2\hbar\omega_q}\right)^2 \left(\frac{e}{C_\Sigma}\right)^2 S_N^0(\omega_q)
	\end{equation}
	\begin{equation}
		\Gamma_\phi  = \frac{1}{4\hbar^2} \left(\frac{e\kappa\varepsilon_0}{\hbar\omega_q}\right)^2 \left(\frac{e}{C_\Sigma}\right)^2 S_N^0(0)
	\end{equation}

	\section{Application to rf-Reflectometry}
	
	\begin{figure*}
		\includegraphics{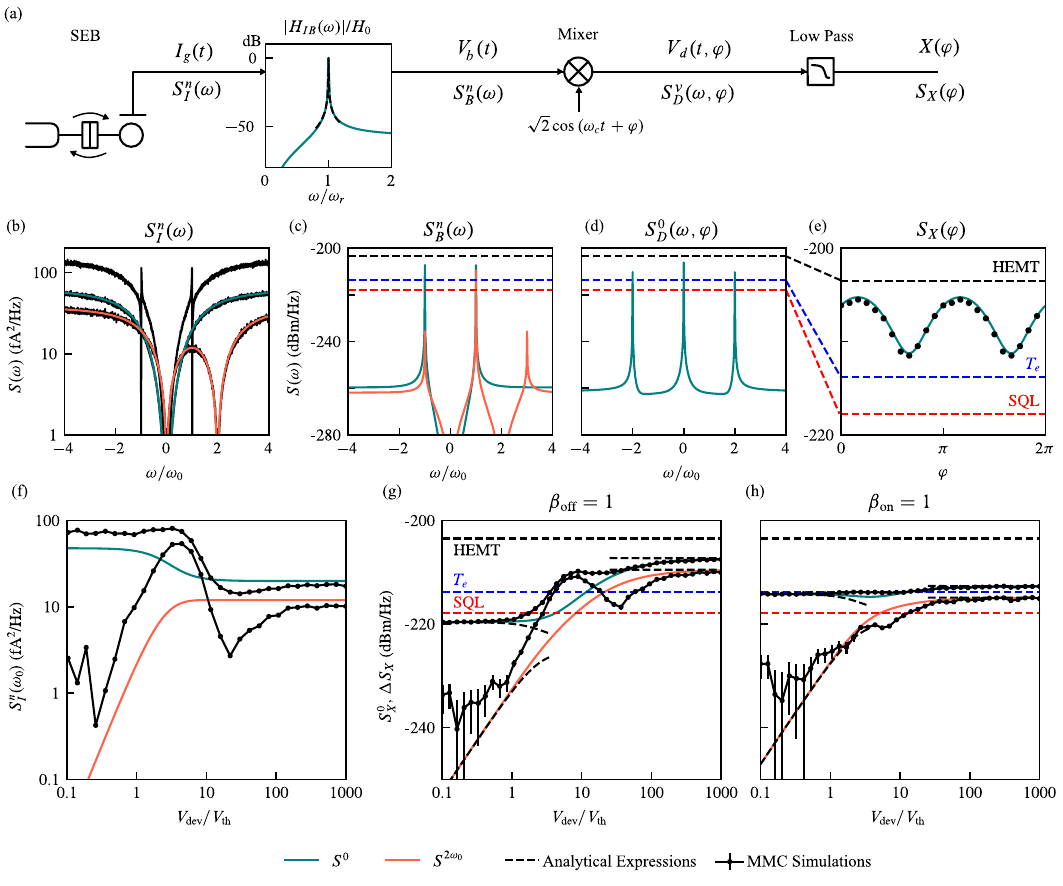}
		\caption{The SEB's intrinsic noise in rf-reflectometry and the efficiency of transmission through the resonator. (a) Signal processing chain associated with the reflectometry circuit in Fig.~\ref{FIG1}(b). The exact (Eq.~\eqref{transfer}, solid line) and approximated (Eq.~\eqref{approx}, dashed line) resonator transfer functions $H_{IB}$ are plotted in the inset, normalised to the peak value $H_0$. (b)-(e) Noise spectra corresponding to each signal in the chain, calculated for typical experimental parameter values (with $\beta_\text{off} = 1$) in the large excitation limit. Black markers are the corresponding MMC simulation results; panel (b) includes spectra in the small (upper line) and large (lower line) excitation limits. Dashed lines correspond to the noise power associated with cryogenic HEMT amplifiers (black), the electron temperature $T_e$ (blue) and the standard quantum limit (SQL- red). (f) Dependence of noise current on excitation amplitude and (g),(h) corresponding demodulated noise power transmitted through two different resonator designs, critically coupled in the unloaded (g) and loaded (h) states respectively. Analytical (dashed), numerical (solid) and MMC simulation (black marker) results are plotted, with HEMT, electron temperature and SQL noise powers included for comparison. The MMC data in panel (f) corresponds to the parameters in panel (g) and is independently calculated from the MMC time-trace $Q(t)$.}
		\label{FIG5}
	\end{figure*}
	
	Having fully characterised the cyclostationary noise spectrum and the intrinsic SNR of the SEB's gate current, we turn our attention to the application of SEB readout via rf-reflectometry, now including the effect of the resonator. The resonator primarily acts as a bandpass filter on the current noise spectrum emitted into the transmission line; however, we shall see that excitation of the resonator by the SEB's intrinsic noise can itself incoherently drive electron tunnelling. This backaction leads to deviations from a pure cyclostationary process, becoming significant at low reflectometry powers (small $V_f$) where the noise excitation on the gate voltage dominates over the coherent sinusoidal drive.
	
	The signal processing chain corresponding to the reflectometry circuit in Fig.~\ref{FIG1}(b) is depicted in Fig.~\ref{FIG5}(a) and consists of bandpass filtering through the resonator with transfer function $H_{IB}(\omega)$ from gate current $I_g$ to transmission line wave $V_b(t)$, followed by synchronous demodulation of $V_b(t)$ as described in section III (downmixing and low pass filtering). In terms of the spectral correlation functions, filtering by the resonator is represented by \cite{gardnerCyclostationarityHalfCentury2006}:
	\begin{equation}
		S_B^{n\omega_0}(\omega) = S_I^{n\omega_0}(\omega) H_{IB}(\omega)H_{IB}^*(\omega-n\omega_0)
	\end{equation}
	where, for a capacitively coupled parallel LCR resonator, 
	\begin{equation}
		H_{IB}(\omega) = \frac{j\omega C_c Z_0}{(Y_Q + Y_\text{res})(1+j\omega C_c Z_0) + j\omega C_c}
		\label{transfer}
	\end{equation}
	using $Y_\text{res} = j\omega C_r + 1/(j\omega L_r) + G_r$ with the lumped element resonator values (capacitance $C_r$, inductance $L_r$, loss conductance $G_r$ and coupling capacitance $C_c$) and transmission line impedance $Z_0$ shown in Fig.~\ref{FIG1}(b). This allows a self-consistent solution which includes the effect of the SEB's small-signal admittance $Y_Q$ loading the resonator. Note that in the large signal limit, power-broadening of the quantum capacitance and Sisyphus resistance means $Y_Q$ tends to zero \cite{derakhshanmamanChargeNoiseOverdrive2020}; for intermediate powers, the reduction of $Y_Q$ can be directly calculated from the fundamental ($\omega_0$) Fourier component of $P_\text{SS}(t)$, or equivalently by using a describing function approach \cite{derakhshanmamanChargeNoiseOverdrive2020,oakesQuantumDotBasedFrequency2023, oakesFastHighFidelitySingleShot2023}.
	
	In Fig.~\ref{FIG5}(b)-(e) we illustrate the noise spectra $S_I^{n\omega_0}$, $S_B^{n\omega_0}$, $S_D^0$ and $S_X$ corresponding to the signals at each stage in the reflectometry chain: gate current $I_g$, reflected wave $V_b$, downmixed signal $V_d$ and low pass filtered output $X$. These are calculated in the large excitation limit and for resonator parameters similar to those seen in recent experimental implementations \cite{ibbersonLargeDispersiveInteraction2021}: resonator frequency $\omega_r = \omega_0 = \omega_c = 2\pi\times \SI{2}{\giga\hertz}$, internal quality factor $Q_i = 2000$, coupling coefficient $\beta = 1$, transmission line impedance $Z_0 = \SI{50}{\ohm}$ and SEB parameters $\Gamma_0/\omega_0 = 1.5$, $T_e = \SI{120}{\milli\kelvin}$, $\alpha = 0.8$ \cite{oakesQuantumDotBasedFrequency2023}. We include for comparison the spectral densities corresponding to the noise temperatures of state-of-the-art cryogenic HEMT amplifiers (\SI{1.3}{\kelvin} \cite{schleehUltralowPowerCryogenicInP2012a, vigneauProbingQuantumDevices2023} - back dashed line) and the electron temperature $T_e$ (\SI{120}{\milli\kelvin} - blue dashed line), as well as the standard quantum limit (SQL) of $\hbar\omega_0/2$ (red dashed line) approached by current parametric amplifier implementations \cite{schaalFastGateBasedReadout2020, muckRadiofrequencyAmplifiersBased2010}. The demodulated noise spectral density $S_X$ [panel~\ref{FIG5}(e)] shows the expected sinusoidal dependence on demodulator phase, that we verify with Markov Monte Carlo simulations (black markers; discussed further below and in appendix C). The spectral density can clearly exceed the standard quantum limit and approach that of the HEMT; this suggests the intrinsic noise of the SEB may directly limit the fidelity of gate-based readout methods in certain resonator and device parameter regimes that we explore next. 

    In particular, we shall consider how the effective noise power in rf-reflectometry depends not only on the SEB's noise current $S_I$, but also on the efficiency by which the noise current is coupled into the transmission line. This is a function of the resonator design, specifically the quality factor $Q_i$ and coupling coefficient $\beta$, as well as the degree of loading of the resonator by the SEB's admittance $Y_Q$. This can be readily seen by considering the approximated form of Eq.~\eqref{transfer},
	\begin{equation}
		H_{IB}(\omega) \approx \frac{\sqrt{KZ_0}}{K+G_r+2j(\omega-\omega_r)C_t + Y_Q}
		\label{approx},
	\end{equation}
	where we define $K = \beta G_r = \omega_r^2C_c^2Z_0 $ and $C_t = C_r + C_c$. For the bare resonator, i.e. with the SEB in Coulomb blockade (`off') or strongly overdriven such that $Y_Q$ tends to zero, the maximum $|H_{IB}|^2/Z_0 = 1/4G_r$ is achieved on resonance ($\omega = \omega_r$) with critical coupling of the bare resonator ($K = G_r$, i.e. $\beta_\text{off} = 1$). With the SEB in the unblockaded, `on' state, loading due to  $Y_Q$ reduces the maximum  to $|H_{IB}|^2/Z_0 = 1/4(G_r + G_\text{Sis})$, achieved on the dispersively shifted resonance ($\omega = \omega_r/(1+C_Q/2C_t)$) and with critical coupling of the loaded resonator ($K = G_r + G_\text{Sis}$, i.e. $\beta_\text{on} = 1$). We see, therefore, that the resonator parameters act in conjunction with the SEB's effective admittance to determine the frequency and excitation amplitude at which the noise is most efficiently coupled.
	
	We demonstrate the above results in Fig. \ref{FIG5}(f)-(h) by calculating, as a function of excitation amplitude, the noise transmitted through two different resonator designs: in \ref{FIG5}(g), for a resonator critically coupled in the unloaded state ($\beta_\text{off} = 1$) and, in \ref{FIG5}(h), for a resonator critically coupled in the loaded state ($\beta_\text{on} = 1$, using a larger value of $C_c$). Though the SEB's gate current noise is, according to our analytical framework, the same for both resonator designs [panel \ref{FIG5}(f), c.f. Fig.~\ref{FIG3}(c)], the noise coupling efficiency into the transmission line, and the excitation amplitude at which this is maximised, differs. For the first resonator ($\beta_\text{off} = 1$), the coupling efficiency is maximised in the strongly overdriven limit when operating at $\omega_0 = \omega_r$, giving a peak value of the average noise of $S_X^\text{max}/Z_0 = S_I^0(\omega_0)/4G_r$, indicated by the horizontal black dashed lines at high $V_\text{dev}$ in Fig.~\ref{FIG5}(g). Conversely, for the second design ($\beta_\text{on} = 1$), the noise is coupled efficiently in the small-signal regime -- although in our example, Sisyphus losses reduce the effective $Q_i$ and hence the efficiency compared to the overdriven regime. In the small-excitation limit with a high-$Q$ resonator ($G_\text{Sis} \gg G_r$), the SEB's intrinsic noise power at the dispersively shifted frequency corresponds to Johnson noise at the reservoir electron temperature: $S_X^\text{max}/Z_0 = k_BT_e/2$ [Fig.~\ref{FIG5}(h)]. We interpolate between the analytical results in the small-signal and large-signal regimes using numerical calculations of both $S_I^{n\omega_0}$ and $Y_Q$ as described above, while tracking the dispersive shift in resonance frequency due to the quantum capacitance with $\omega_0$ [green and orange solid lines in Fig.~\ref{FIG5}(g)-(h)]. However, as we discuss next, our autocorrelation-based model cannot fully capture the dynamics of the hybrid system: the effective admittance $Y_Q$ in the nonlinear regime is only valid at the excitation frequency and cannot represent the response to wideband noise, while our analysis assumes the gate voltage $V_g$ to be a noiseless sinusoid.
	
	To improve upon our above calculations and investigate the effect of noise-driven tunnelling, we resort to MMC simulations of the complete circuit including the SEB, the resonator and synchronous demodulation. The ring-down time of the resonator induces memory into the combined system, as captured by its step response to each electron tunnelling event.  Augmenting the model in Eq.~\eqref{MMC} with the non-Markovian dynamics of the resonator can then be represented by the convolution
 	\begin{equation}
 		\begin{split}
 		V_{g(b)}(t) = &\Re\left(H_{FG(FB)}(\omega_0)V_fe^{j\omega_0t}\right)\\
 		              &+ \int_{-\infty}^{t}I_g(\tau)h_{IG(IB)}(t-\tau)\, d\tau
 		\end{split}
 		\label{nonmarkov}
 	\end{equation}
	where $H_{FG(FB)}(\omega)$ is the resonator transmission (reflection) coefficient $V_{g(b)}/V_f$, and $h_{IG(IB)}(t)$ is the gate current impulse response (i.e. charge step response) of $V_{g(b)}(t)$; we implement the convolution using an infinite impulse response digital filter (details in appendix C). $S_X^0$ and $\Delta S_X$ [black markers in Fig.~\ref{FIG5}(g)-(h)] are extracted from the phase-dependent $S_X(\varphi)$ (c.f. Fig~\ref{FIG5}(e)), with the mean and standard error calculated from 500 trials of duration $2^{14}$ cycles with a step size $\Delta t = T/2^{11}$. 
	
	The simulations show good agreement with the master equation analysis when the resonator is strongly coupled to the transmission line (large $C_c$ as in Fig~\ref{FIG5}(h)) or in the limit of large excitation powers. However, for weaker resonator-transmission line coupling, as is seen in Fig~\ref{FIG5}(g), significant discrepancies occur at low and intermediate excitation powers; in this example, we observe increases in both the average noise and the degree of cyclostationary correlation around $V_\text{dev}/V_\text{th} = 10$. This suggests that the non-Markovian dynamics and incoherent excitation of the resonator by the SEB, which are not captured in our autocorrelation analysis, can no longer be neglected when the resonator response to individual tunnelling events is large compared to the periodic drive. The discrepancy can be traced back to the current noise emitted by the SEB [black markers in Fig.~\ref{FIG5}(f)], which we compute directly from the MMC time series $Q(t)$ (appendix C). The corresponding spectrum for the zero-excitation limit [upper black line in Fig.~\ref{FIG5}(b)] demonstrates a sharp perturbation around the frequency of the resonator, but conforms to the analytical results in the large excitation limit [lower black line in Fig.~\ref{FIG5}(b)]. The noise peak seen at intermediate powers in Fig.~\ref{FIG5}(f)-(g) can be explained by the operating frequency $\omega_0$ sweeping across this perturbation around $\omega_r$ as we track the dispersive shift. However, for a deeper understanding of the interactions occurring in the hybrid SEB-resonator system, a full analysis of the underlying stochastic differential equations would be necessary.
	
	\section{Conclusions}

    We have addressed the lack of a noise model applicable to radio frequency readout of the Single-Electron Box, an increasingly widespread technology in the development of semiconductor quantum computing architectures. Starting from a semiclassical model of stochastic electron tunnelling between a quantum dot and a reservoir driven by a periodic gate excitation, we described the noise mechanism as a cyclostationary processes to derive and develop analytical expressions, numerical models and Markov Monte Carlo simulations for the gate current noise. Our analytical framework is able to precisely capture the correlations that arise in the noise spectrum and the interference that results upon demodulation or lock-in detection of the SEB signal, which manifests as phase-dependent noise enhancement or suppression.

    In addition to characterising the SEB's gate noise current, we have studied its practical application to qubit readout by rf-reflectometry, considering the dynamics of the combined SEB-resonator system and evaluating the noise power under typical experimental parameters. The resonator design determines the efficiency by which the SEB's noise current is emitted into the transmission line; with the resonator critically coupled to the line ($\beta = 1$), we find the emitted noise power can approach that of cryogenic HEMT amplifiers. Conversely, under weak resonator-line coupling and small rf drive powers, excitation of the resonator by the SEB's intrinsic noise can itself incoherently drive electron tunnelling. Rich interaction dynamics within the hybrid SEB-resonator system emerge, which we can probe with Markov Monte Carlo simulations but are beyond the capacity of our autocorrelation-based analytical approach. 
    
    With the introduction of parametric amplifier-enhanced QD readout hardware, we expect signatures of the intrinsic noise of the SEB to be measured in the near future. This will allow the experimental realisation of schemes to achieve a high degree of noise suppression by exploiting spectral correlations and enhance the fidelity spin qubit readout with SEBs.

	\begin{acknowledgments}
		The authors acknowledge helpful discussions with Lorenzo Peri and Luca Gammaitoni. L.C. acknowledges support from EPSRC Cambridge UP-CDT EP/L016567/1 and M.F.G.Z. acknowledges support from the European Union’s grant agreement No. 951852, Innovate UK Industry Strategy Challenge Fund (10000965) and the UKRI Future Leaders Fellowship Programme (MR/V023284/1).
	\end{acknowledgments}
	
	\appendix
	
	\section{Derivation of the Cyclostationary Spectral Correlation Functions}
		Under the transformation $\tau = t - t_0$, the noise autocorrelation can be expressed as
		\begin{equation}
			R_N(t, \tau) = R_Q(t,\tau) - P_\text{SS}(t)P_\text{SS}(t+\tau)
			\label{RNtt}
		\end{equation}
		with
		\begin{equation}
			\begin{split}
				R_Q(t,\tau) = &h(\tau)P_\text{SS}(t)P_1(t +\tau, t) + \\
				              &h(-\tau)P_\text{SS}(t+\tau)P_1(t, t+\tau)
			\end{split}
		\end{equation}
		where $h(\tau)$ is the Heaviside step function. Inserting Eq.~\eqref{Eq:master_sol}, the solution to the linearised master equation, into \eqref{RNtt}, 
		\begin{equation}
			\begin{split}
			R(t,\tau) = &h(\tau)e^{-\Gamma_0\tau}P_\text{SS}(t)\left(1-P_\text{SS}(t)\right) + \\ &h(-\tau)e^{\Gamma_0\tau}P_\text{SS}(t+\tau)\left(1-P_\text{SS}(t+\tau)\right)
			\end{split}
			\label{Rtau_expanded}
		\end{equation}
		and equating terms with Eq.~\eqref{Fourier}, we find the Fourier decomposition of $R_N(t,\tau)$ can be written as:
		\begin{equation}
			R^0_N(\tau)  =  \left[\frac{1}{4\cosh^2(V_0/2V_\text{th})}- \frac{1}{2}\left(\frac{V_\text{dev}|Y_Q|}{e\alpha\omega_0}\right)^2  \right] e^{-\Gamma_0 |\tau|}
		\end{equation}
		\begin{equation}
			\begin{split}
				R_N^{\omega_0}(\tau) = &\frac{V_\text{dev}Y_Q}{e\alpha\omega_0} \left(\frac{1}{2}-P_\text{th}\right) \\  
				&\times \left[ h(\tau)e^{-\Gamma_0\tau} +  h(-\tau)e^{\Gamma_0\tau}e^{j\omega \tau} \right]
			\end{split}
		\end{equation}
		\begin{equation}
			R_N^{2\omega_0}(\tau) = -\left(\frac{V_\text{dev}Y_Q}{2e\alpha\omega_0}\right)^2 \left[h(\tau)e^{-\Gamma_0\tau} + h(-\tau)e^{\Gamma_0\tau}e^{2j\omega\tau} \right] 
		\end{equation}
		with $R_N^{-\omega_0}(\tau)$ and $R_N^{-2\omega_0}(\tau)$ given by the complex conjugates of $R_N^{\omega_0}(\tau)$ and $R_N^{2\omega_0}(\tau)$. Taking the Fourier Transform of the above cyclic autocorrelation functions gives the set of charge noise spectral correlation functions:
		\begin{equation}
			S_N^0(\omega) = \left[\frac{1}{4\cosh^2(V_0/2V_\text{th})}- \frac{1}{2}\left(\frac{V_\text{dev}|Y_Q|}{e\alpha\omega_0}\right)^2  \right] \frac{2\Gamma_0}{\Gamma_0^2 + \omega^2}
		\end{equation}
		\begin{equation}
			\begin{split}
			S_N^{\omega_0}(\omega) = &\frac{V_\text{dev}Y_Q}{e\alpha\omega_0}\left(\frac{1}{2}-P_\text{th}\right) \\ &\left[\frac{1}{\Gamma_0 + j\omega} + \frac{1}{\Gamma_0 - j(\omega-\omega_0)}\right]
			\end{split}
		\end{equation}
		\begin{equation}
			S_N^{2\omega_0} = -\left(\frac{V_\text{dev}Y_Q}{2e\alpha\omega_0}\right)^2 \left[\frac{1}{\Gamma_0 + j\omega} + \frac{1}{\Gamma_0 - j(\omega-2\omega_0)}\right]
		\end{equation}
		
		Similarly, inserting \eqref{Pss_Square} into \eqref{Rtau_expanded} and considering the Fourier series of the resulting waveform,
		\begin{equation}
			P_\text{SS}(t)\left(1-P_\text{SS}(t)\right) = \beta e^{-\Gamma_0(t + T/4)}\left(1-\beta e^{-\Gamma_0(t + T/4)}\right)
		\end{equation}
		for $-T/4\leq t<T/4$, we derive the results for large excitation limit:
		\begin{equation}
			R^0_N(\tau) = \frac{\omega_0}{2\pi\Gamma_0}\tanh\left(\frac{\pi\Gamma_0}{2\omega_0}\right)e^{-\Gamma_0|\tau|}
		\end{equation}
		\begin{equation}
			\begin{split}
			R_N^{2\omega_0}(\tau) = &\frac{\omega_0}{\pi}\tanh\left(\frac{\pi\Gamma_0}{2\omega_0}\right)\left[\frac{1}{\Gamma_0+2j\omega_0} - \frac{1}{2(\Gamma_0+j\omega_0)}\right]\\
			&\times\left[h(\tau)e^{-\Gamma_0\tau} + h(-\tau)e^{\Gamma_0\tau}e^{2j\omega\tau} \right]
			\end{split}
		\end{equation}
		\begin{equation}
			S^0_N(\omega) = \frac{\omega_0}{\pi}\tanh\left(\frac{\pi\Gamma_0}{2\omega_0}\right)\frac{1}{\Gamma_0^2 + \omega^2}
		\end{equation}
		\begin{equation}
			\begin{split}
			S_N^{2\omega_0}(\omega) = &\frac{\omega_0}{\pi}\tanh\left(\frac{\pi\Gamma_0}{2\omega_0}\right)\left[\frac{1}{\Gamma_0+2j\omega_0} - \frac{1}{2(\Gamma_0+j\omega_0)}\right]\\
			&\times\left[\frac{1}{\Gamma_0 + j\omega} + \frac{1}{\Gamma_0 - j(\omega-2\omega_0)}\right]
			\end{split}
		\end{equation}
		Note that due to the symmetry of $P_\text{SS}(t)$, the first order terms $R_N^{\omega_0}(\tau)$ and $S_N^{\omega_0}(\omega)$ are zero.
 	
		\section{Backaction}
		
		We are interested in the time-averaged autocorrelation of the voltage fluctuations $V_\text{QD}(t)$ on the SEB's quantum dot, $R_V^0(\tau)$, which can be calculated from
		\begin{equation}
			\begin{split}
				&R_V(t_1,t_2) = \mathbb{E} \left[V_\text{QD}(t_1) V_\text{QD}(t_2)\right]\\
				&= \frac{e^2}{C_\Sigma^2}\mathbb{E} \left[Q(t_1) Q(t_2)\right] + \alpha^2V_g(t_1)V_g(t_2)\\
				&- \frac{e\alpha}{C_\Sigma} \Big[\mathbb{E} \left[Q(t_1)\right]V_g(t_2) + \mathbb{E} \left[Q(t_2)\right]V_g(t_1)\Big]. \\
			\end{split}
		\end{equation}
		Recognising that $\mathbb{E} \left[Q(t_1) Q(t_2)\right] = R_Q(t_1, t_2)$ and $\mathbb{E} \left[Q(t)\right] = P_\text{SS}(t)$, we can write down
        \begin{equation}
            \begin{split}
                R_V(t_1,t_2) = &\frac{e^2}{C_\Sigma^2}R_{N}(t_1,t_2) + \frac{e^2}{C_\Sigma^2} P_\text{SS}(t_1)P_\text{SS}(t_2) \\ 
		          &- \frac{e\alpha}{C_\Sigma}\Big[P_\text{SS}(t_1)V_g(t_2) + P_\text{SS}(t_2)V_g(t_1)\Big] \\
                &+ \alpha^2V_g(t_1)V_g(t_2)
            \end{split}
        \end{equation}
        The first term gives the time-averaged voltage noise associated with charge fluctuations $S_V(\omega)$ (from the Fourier transform of $R_V^0(\tau) = \frac{e^2}{C_\Sigma^2} R_N^0(\tau)$). Evaluating the complete expression in the small-signal limit gives:
        \begin{equation}
			\begin{split}
				R_V^0(\tau) = 
                & \frac{e^2}{C_\Sigma^2} R_N^0(\tau) + \frac{e^2}{C_\Sigma^2}P_\text{th}^2 + 
				\frac{\cos\omega_0\tau}{2} \bigg[ \alpha^2V_\text{dev}^2 \\
				& + \frac{e^2}{C_\Sigma^2}\left(\frac{V_\text{dev}|Y_Q|}{e\alpha\omega_0}\right)^2 + \frac{2e\alpha}{C_\Sigma}\left(\frac{V_\text{dev}^2G_\text{Sis}}{e\alpha\omega_0}\right) \bigg].
			\end{split}
		\end{equation}
		where the remaining terms can be identified as due to the harmonic component of the QD voltage
		\begin{equation}
			V_\text{QD}^H(t) = \frac{eP_\text{th}}{C_\Sigma}+\Re\left\{\left(\frac{Y_Q}{\alpha j\omega_0C_\Sigma} + \alpha\right)V_\text{dev} e^{j\omega_0 t}\right\}.
		\end{equation}
		Note that additional terms at frequencies $n\omega_0$ would arise in  $V_\text{QD}^H(t)$ in the large-excitation due to harmonic distortion in the exact solution to the nonlinear master equation.
		
		\section{Markov Monte Carlo Simulations}
		
		We discretize the time-inhomogeneous Markov process introduced in Eq.~\eqref{MMC} at $N_\Delta$ samples per cycle period $T = 2\pi/\omega_0$, with successive samples $Q_k = Q(kT/N_\Delta)$ calculated from
		\begin{equation}
			Q_{k+1} = R^+_k + (1-R^+_k-R^-_k)Q_k
		\end{equation}
		where the tunnelling in (out) transition flags $R^{+(-)} \in \{0,1\}$ are Bernoulli random variables set by comparison to a uniform random sample $U_k \sim \mathcal{U}(0,1)$ according to
		\begin{equation}
			R^{+(-)}_k = 1 \quad \text{if} \quad \frac{T}{N_\Delta} \Gamma_\text{in(out)}(V_{g,k}) < U_k.
		\end{equation}
		
		Here, $V_{g,k}$ includes both the harmonic resonator response due to the excitation signal and the response $V_g(t)$ due to the tunnelling events in the SEB (steps in $Q(t)$), represented by the convolution in Eq. \eqref{nonmarkov} of the main text. The relevant transfer functions can be derived from simple circuit theory, using the values defined in the main text, as 
		\begin{equation}
			\begin{split}
				H_{QG}(s) &= \frac{V_g(s)}{Q(s)}= \frac{\alpha s^2 L (1+s C_c Z_0)}{B + s^2 C_c L} \\
				H_{QB}(s) &= \frac{V_b(s)}{Q(s)}=\frac{\alpha Z_0 s^3 LC_c}{B + s^2 C_c L}\\
				H_{FG}(s) &= \frac{V_g(s)}{V_f(s)}=\frac{2}{1+(sC_r+G_r+1/sL)(1/sC_c+Z_0)}\\
				H_{FB}(s) &= \frac{V_b(s)}{V_f(s)}=\frac{1+(sC_r+G_r+1/sL)(1/sC_c-Z_0))}{1+(sC_r+G_r+1/sL)(1/sC_c+Z_0)}
			\end{split}
		\end{equation}
		where $B = (1+s C_c Z_0) (1+sLG_r +  s^2 (C_r+\alpha C_S) L)$.
		As $Q(t)$ is composed only of steps, the convolution can be implemented exactly by an infinite impulse response (IIR) discrete-time filter with coefficients calculated by a step-invariant transformation of the continuous-time resonator transfer function:  
		\begin{equation}
			H_\text{IIR}(z) = \left(\frac{z-1}{z}\right) \mathcal{Z}\left(\mathcal{L}^{-1}\left(\frac{H_{QG(QB)}(s)}{s}\right)_{t=kT/N_\Delta}\right).
		\end{equation}
		After sampling the inverse Laplace transform of the partial fraction expansion of $H(s)/s$ at a rate $\omega_0N_\Delta/2\pi = N_\Delta/T$
		\begin{equation}
			\mathcal{L}^{-1}\left( \sum_i \frac{K_i}{s+s_i} \right)_{t=kT/N_\Delta} 
			= \sum_i K_i e^{-s_ikT/N_\Delta},
		\end{equation}
		where $i = \{0, 1, \dots , n_p-1\}$ and $n_p$ is the number of poles of $H(s)/s$, taking the Z-transform gives the set of digital filter coefficients $\left\{a_i\right\}, \left\{b_i\right\}$ according to:
		\begin{equation}
			\sum_i \frac{K_i}{1+e^{-(s_iT/N_\Delta)z^{-1}}} = \frac{\sum_i b_i z^{-i}}{\sum_i a_i z^{-i}}.
		\end{equation}
		We implement this filter in Direct Form II, with $m$ being an internal state of the filter:
		\begin{equation}
			m_k = -|e|Q_k - \sum_i a_i m_{k-i} \, ; \quad
			V_k = \sum_i b_i m_{k-i}.
		\end{equation}		
		
		Synchronous demodulation is modelled numerically by downmixing $V_d(t) = V_b(t) \times \sqrt{2}\cos{(\omega_c t + \varphi)}$ followed by a digital low pass filter ($5^\text{th}$-order Butterworth with cut-off frequency $\omega_0/10$). The sampling rate of the output $V_X(t)$ is then decimated from $N_\Delta \omega_0/(2\pi)$ to $\omega_0/(2\pi)$, accelerating the power spectral density estimation without aliasing or loss of information around zero frequency, and the resonator transient response (of duration $5Q_L$ cycles, covering 99.3\% of the transient decay) is cropped.
		
		\begin{figure}[t]
			\includegraphics{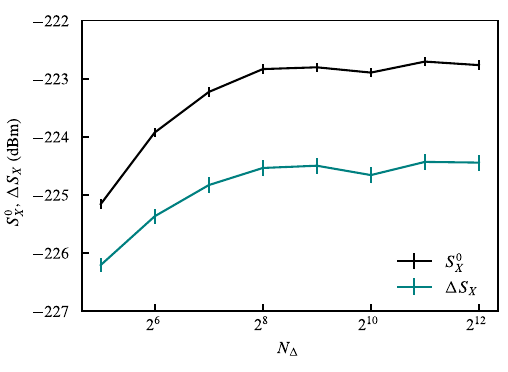}
			\caption{Convergence of simulated spectral correlation densities with $N_\Delta$.}
			\label{fig:convergence}
		\end{figure} 
	
		We estimate the zero-frequency noise power spectrum $S_X(0)$ from an FFT-based periodogram calculation, taking the component at the lowest non-zero frequency $\omega_0/{N_\text{cyc}}$ where $N_\text{cyc}$ is the duration of the simulation, with the standard error calculated from the variance of $N$ trials. Simulations demodulated at different local oscillator phases $\theta$ are fitted to the trochoid function $S_X(\varphi) = S_X^0 + \Delta S_X \cos(2\varphi + \theta)$ (as introduced in the main text; see also figure \ref{FIG5}(e)) gives the final estimates of the cyclostationary spectral correlation densities  $S_X$  and $\Delta S_X$, with errors calculated from the covariances of the fitting parameters.
		
		In the case of the current noise spectral correlation functions $S_I^{n\omega_0}(\omega)$, the estimate is derived directly from the time series $Q(t)$ using the relation \cite{gardnerCyclostationarityHalfCentury2006}:
		\begin{equation}
			R^{n\omega_0}_N(\tau) = \int_{-\infty}^{\infty}N(t)N(t+\tau)e^{jn\omega_0 t} \,dt
		\end{equation}
		in conjunction with Eq.~\eqref{NtoI}.

		We pick the simulation time step by assessing the convergence of the extracted spectral correlation densities with increasing $N_\Delta$ as shown in Figure \ref{fig:convergence}; for 500 trials of duration $N_\text{cyc} = 2^{14}$, the values converge to within the standard error by $N_\Delta = 2^{11}$.
		
    %
	
\end{document}